**Development of Automatic Speech Recognition for Kazakh Language using Transfer Learning**

Amirgaliyev E.N., Kuanyshbay D.N., Baimuratov O.

**Abstract**

Development of ASR (Automatic Speech Recognition) system for kazakh language is very challenging due to a lack of data. Existing data of kazakh speech with its corresponding transcriptions are heavily accessed and not enough to gain a worth mentioning results. For this reason, speech recognition of kazakh language hasn't been explored well. There are only few works that investigate this area with traditional methods (Hidden Markov Model, Gaussian Mixture Model), but they are suffering from poor outcome and lack of enough data. In our work we suggest a new method that takes pre-trained model of russian language and applies its knowledge as a starting point to our neural network structure, which means that we are transferring the weights of pre-trained model to our neural network. The main reason we chose russian model is that pronunciation of kazakh and russian languages are quite similar because they share 78% letters and there are quite large corpus of russian speech dataset. We have collected a dataset of kazakh speech with transcriptions in the base of Suleyman Demirel University with 50 native speakers each having around 400 sentences. Data have been chosen from famous kazakh books like "Abay zholy", "Kara sozder" etc.

We have considered 4 different scenarios in our experiment. First, we trained our neural network without using a pre-trained russian model with 2 LSTM (Long-Short-Term Memory) layers and 2 BiLSTM (Bidirectional Long-Short-Term Memory). Second, we have trained the same 2 LSTM layered and 2 BiLSTM layered using a pre-trained model. As a result, we have improved our model's training cost and Label Error Rate (LER) by using external Russian speech recognition model up to 24% and 32% respectively. Pre-trained Russian language model has trained on 100 hours of data with the same neural network architecture.

**Introduction**

Automatic Speech Recognition tasks are very challenging, although the results are improving and growing due to the raise of required data, advancement of graphical processors (GPU) and "fine-tuning" of neural network. The most accurate and state-of-the-art ASR systems [1-4] for English language and mandarin that has been developed recently uses Switchboard, Fisher, TIMIT dataset with over 2000 hours of continuous speech data. This amount of data is clearly enough to set a particular experiment and gain promising results. Despite the usage of recent neural network techniques the main influence of desired output is the dataset capacity, which for popular languages like English is growing fast. Another important influence on accurate ASR system is the selection of right neural network structure and tuning parameters, which leads to fast training and resolving the overfitting problem.

End-to-end ASR systems have already overcome the traditional HMM and DNN systems due to its simplicity and convenience, where there is no need to have the usage of language model, pronunciation model etc. This [5] model have been built with the help of the technique, which is called CTC (Connectionist Temporal Classifier). CTC makes the automatic segmentation of audio signal and maps the audio wave directly to transcriptions. Neural network structure based on RNN (recurrent neural network) where, each neuron returns the probability distribution of all characters including the blank space for each segment of an audio wave (Figure 1). To find the CTC loss they sum all corresponding sequence distributions. Decoding part is done by applying an algorithm called beam search or greedy search.

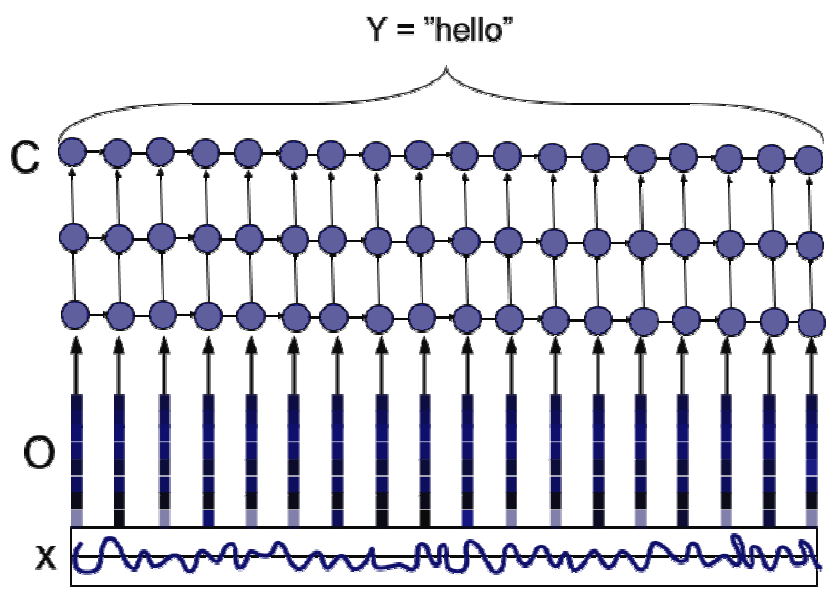

Figure 1. Connectionist Temporal Classifier

The model that has been developed here [6] uses the same technique with a small advancement, which is called attention based CTC model. Basically, they combine the CTC loss function with attention function that is used in sequence-to-sequence model to build an effective and robust ASR system. This method improved the accuracy of the simple CTC model up to 17%, which was a huge leap forward.

Recent usages of deeper LSTM networks show that they noticeably outperform the fully connected neural networks [14]. The approach that was presented here [12] using LSTM on speech recognition task reduces the character error rate down (CER) to 14% on Shenma voice search data. They were able to train a deep 7 layer LSTM network with using a layer-wise training and exponential moving average methods. In this paper [13] authors present the projection layer between output layer and LSTM layer. The presented method improves the performance of this LSTM over the traditional LSTM.

The main disadvantage of these models [5-6], they require a large amount of data, which is a big problem for Kazakh language, because it hasn't been investigated and researched well. The datasets that exist today in Kazakh language are mostly private and not available for free. Even so, these datasets are not big enough to get a good result.

Considering all the disadvantages and obstacles, we have partially solved the problem with a lack of data. Firstly, we have built a convenient website tool that contains a lot of sentences in Kazakh language and allows the user to pronounce the record and save these sentences in a comfortable way. Such a way, we have collected almost 20 hours of data and it continues to grow. Secondly, we have taken the existing [7] speech recognition model for Russian language based on VoxForge dataset with over 100 hour of speech data with corresponding text transcriptions and copied the weights of first 2 LSTM layers with 128 neurons and pasted to our neural network with the same amount of layers and neurons. We didn't take the last layer weights because it's size doesn't match with our layer's size, since the number of characters in Russian alphabet is less than in Kazakh alphabet. As a result, we have lowered our LER down to 19% with the help of external Russian language model, which is a great leap forward.

The paper is organized as follows: Section 2 explains the transfer learning in speech recognition systems. Section 3 discusses about datasets and preprocessing that has been done. Section 4 contains an experiment with 4 scenarios with its construction illustrations and the results. Section 5 reveals the future plans related to ASR system of Kazakh language and data collection methods. Section65 makes the conclusion of the experiment.

**Transfer learning**

Transfer learning is a novel approach that makes the training a lot better and accurate by transferring the knowledge from different task to a current task (Figure 2).

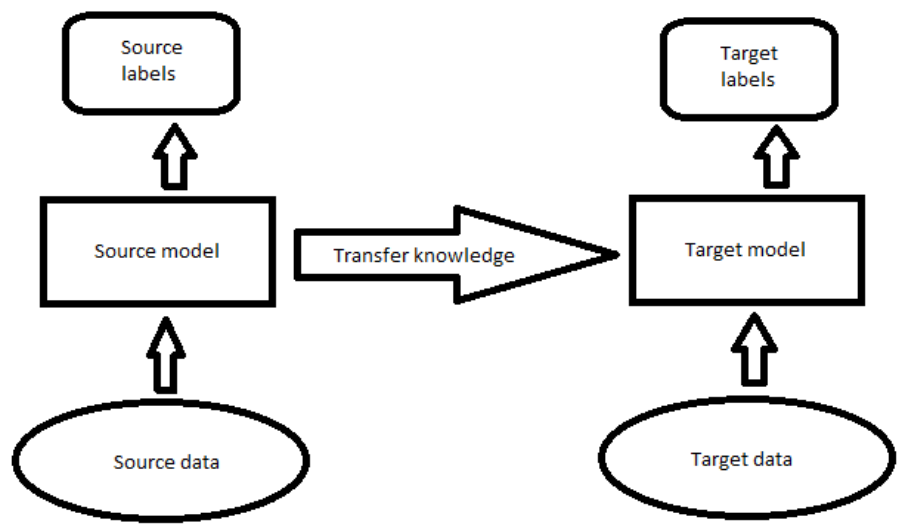

**Figure 2. Transfer learning illustration**

Just as human beings, any model can be trained, learned significantly faster and efficient, if it had a previous experience on related tasks. Usually, source model is the model that has been trained on a massive amount of data, whereas the target model can be trained on small amount. Therefore, transfer learning can solve a problem with a lack of data. Nowadays, there are a lot of pre-trained models that are open-sourced and can be accessed very easily. The main idea behind transfer learning is to transfer the features and parameters (weights) from source task to target task. Basically, source model is considered as a starting point for a target model.

There are a lot of researches being done on multilingual speech recognition tasks [15-19]. The most common problem that they are facing till these days is that languages are specific. Language adaptive acoustic models can be built if languages share acoustic, pronunciation, phonetic properties.

This [20] paper presents the interesting approach on different region specific Indian 9 Languages, where each of these languages share the same phonetic, acoustic features and properties. They have combined all grapheme sets and trained on a sequence-to-sequence model. The result they have gained was 21% improvement on performance compared to the same model trained individually for each language.

For English, German and Spanish data from Cortana there has been an experiment presented here [21]. They have built a multilingual ASR system by using the universal character set shared around all languages. Each language had 150 hours of training set and 10 hours of validation and test set. By using [22], they have resulted 81 English labels, 93 German label and 97 Spanish labels. Therefore, they have built a universal label set (108) with 75% shared overlapping labels. By creating the mechanism which is called the language specific gating mechanism they have trained their model, which outperforms the monolingual approach.

In this [8] paper they applied the transfer learning to a Text-to-Speech task, which is able to generate an audio with data, that has never been seen before. They have uses 3 different pre-trained components: 1) *speaker encoder network* that has been trained on noisy speech dataset with no transcriptions; 2) *synthesis network* which is trained to generate mel spectrogram to text; 3) *vocoder network* which converts mel spectrogram to the waveform with time domain.

This [9] paper applies the transfer learning approach on developing ASR system for German language. With a limited training data, they have adapted a Wav2Letter model, which is originally trained on English language. The paper [10] presented by Vu and Schultz have developed multilingual Multilayer Perceptrons (MLP). This MLP later was applied as a starting point on target languages like Vietnamese, Czech and Hausa. As a result, they have obtained the improvement on error rate up to 22%.

The approach that we take is almost the same as [9, 10], in which we are developing an ASR system for Kazakh language with the help of Russian language. The pre-trained model for Russian language has been trained on VoxForge dataset having a neural network structure with 2 LSTM layers with 128 neurons each and dense layer. Neural network has been trained with 700 epochs and has the same model with Bidirectional LSTM. It uses the same CTC loss function (Figure 3).

Basically, we are using a pre-trained model to extract the weight matrix and copy to our exactly the same neural network structure with 2 LSTM layers. After application of "fine-tuning" our neural network, we start to train 20 hour of Kazakh speech dataset. Simply, we are initializing the weights of our neural network.

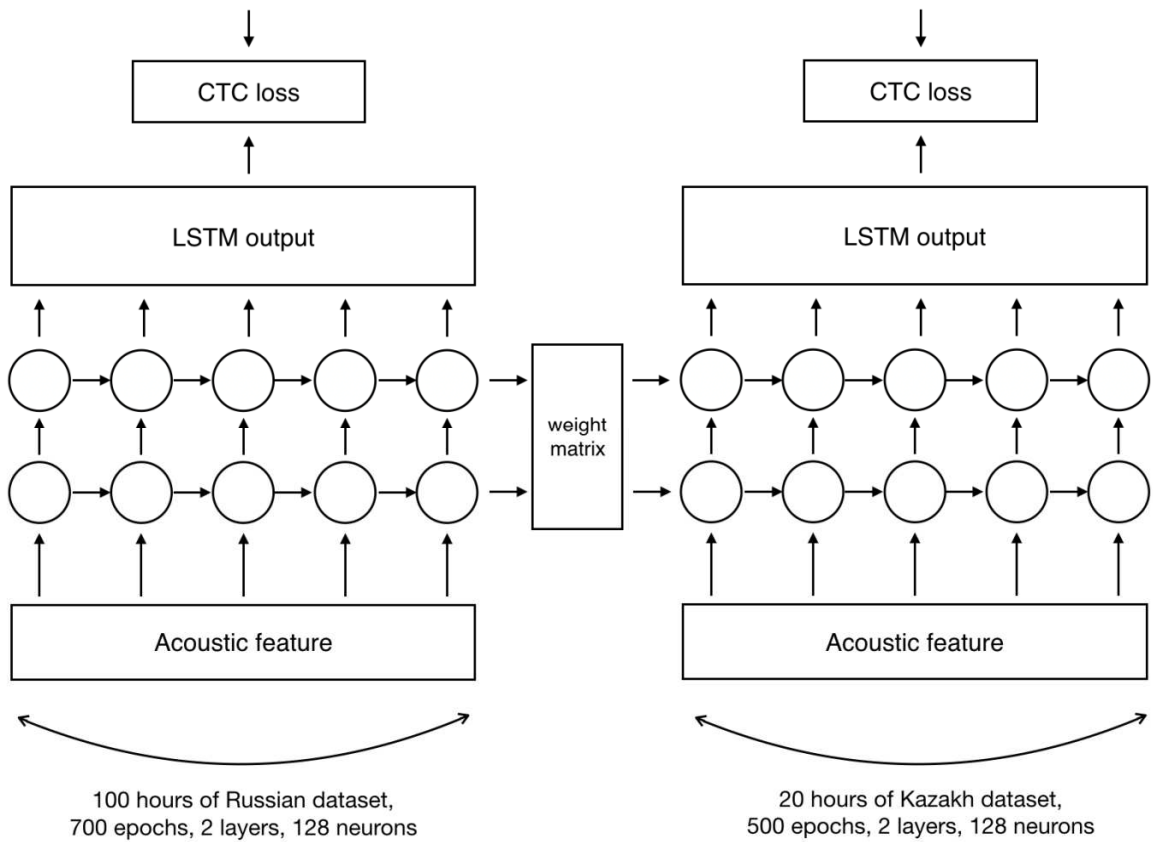

**Figure 3. Weight transfering**

## Datasets and preprocessing

The dataset for Kazakh language has been collected in the base of Suleyman Demirel University. By using special tool (website), 50 native speakers have been involved in pronouncing and saving their utterances. The sentences have been collected from famous Kazakh books "Kara sozder", "Abay zholy" etc. Each speaker has pronounced approximately 400 sentences. Audio files with duration longer than 15 seconds have been omitted, in order to have a strong dependency between transcriptions and audio files. Generally, we have gained around 20 hours of data. All audio files have been sampled to 16 kHz. Using a librosa library, feature extraction has been done by the Mel Frequency Cepstral Coefficients (MFCC). Text files have been normalized by removing all the unnecessary characters and representing in lower case.

The overall duration of our speech corpus, slightly inferior compared to corpus that has been described here [11], where they have biggest speech corpus for Kazakh language. They have collected around 30 hours of data with 200 different speakers with different genders and ages. Since, our dataset has been collected in limited amount of time with special developed website; we will soon pass their corpus in term of duration.

**Experiments and results**

For our network we have built RNN based architecture with 2 Long-Short Term Memory layers and 1 dense layer. The training has been done on several Graphical Processors (GPU) Tesla K80. The environment that has been selected for this task is Jupiter Notebook on Python language (Tensorflow library). Dataset has been cloned from github repository and split 80%, 10% and 10% for training, validation and testing respectively. The parameters of neural network are the following:

- 2 layers of LSTM and BiLSTM separately
- 128 neuron for each layer
- 500 epochs
- Dropout layer after each LSTM layer with 50% probability
- Batch size is 4
- Momentum value for MomentumOptimizer is 0.9
- Learning rate is 0.0005
- CTC loss function
- Metric is Label Error rate (LER)

**Table 1. Results of training**

| RNN type | Training cost | Training LER | Validation cost | Validation LER | Epochs |
|---|---|---|---|---|---|
| LSTM | 15.603327 | 0.054680485 | 3.98756 | 0.01569438 | 500 |
| LSTM with Russian model | 14.426534 | 0.056311063 | 4.78974 | 0.01453637 | 500 |
| BiLSTM | 18.366533 | 0.062855324 | 4.25945 | 0.01602836 | 500 |
| BiLSTM with Russian model | 13.924501 | 0.042720266 | 3.87567 | 0.014177615 | 500 |

We have considered 4 different scenarios (Table 1): 1) LSTM neural network without Russian model; 2) LSTM neural network with Russian model; 3) Bidirectional LSTM without Russian model; 4) Bidirectional LSTM with Russian model. The result of each recurrent neural network actually very close, but we see that the architectures with transfer learning clearly make an improvements on everything.

LSTM layered neural network with external model has improved the training cost up to 8%, whereas Label error rate has increased up 4%. Bidirectional LSTM has showed very promising results, improving the training cost up to 24% and decreasing the label error rate down to 32%.

The experiments above showed that using an external Russian ASR system model to transfer its knowledge to Kazakh language system improves the performance decently.

**Future plans**

For future works, we are planning to accomplish the following tasks:

1) Collect more data in Kazakh language using our website tool
2) Integrate a language model in order to improve the performance using text corpora
3) Integrate other external models that was trained on even bigger data
4) Try to use different neural networks for comparison purposes
5) Compare the results with different audio features other than MFCC

**Conclusion**

We have concluded that using an external Russian language model can partially solve the lack of data problem. We have trained our model using 2 different neural networks (LSTM, BiLSTM) and each of them trained by transferring the weights from external model. This external model was trained on VoxForge dataset with 100 hours of Russian speech. For our own dataset, we have collected around 20 hours of Kazakh speech using famous Kazakh books. Results showed that BiLSTM model with external Russian model improved the performance very well, lowering the training cost and LER down to 24% and 32% respectively.